\documentclass[12pt,a4paper]{article}
\usepackage{fullpage}
\usepackage{cite}
\usepackage{amsbsy}
\usepackage{amsfonts}
\usepackage{amssymb}
\usepackage[dvips,monochrome]{color}
\usepackage[dvips]{graphicx}
\begin{document}
\begin{titlepage}
\begin{flushright}
\begin{tabular}{l}
hep-ph/0202063
\\
7 February 2002%\today
\end{tabular}
\end{flushright}
\vspace{1cm}
\begin{center}
\large\bfseries
The Phase of Neutrino Oscillations
\\[0.5cm]
\normalsize\normalfont
C. Giunti
\\
\small\itshape
INFN, Sezione di Torino,
\\
\small\itshape
and
\\
\small\itshape
Dipartimento di Fisica Teorica,
Universit\`a di Torino,
\\
\small\itshape
Via P. Giuria 1, I--10125 Torino, Italy
\end{center}
\begin{abstract}
Using an analogy with the well-known double-slit experiment,
we show that the standard phase of neutrino oscillations
is correct,
refuting recent claims of a factor of two correction.
We also improve the wave packet treatment of neutrino oscillations
taking into account explicitly
the finite coherence time of the detection process.
\end{abstract}
\end{titlepage}

The experimental and theoretical investigation of neutrino oscillations
is presently one of the most active fields of research
in high-energy physics.
For the interpretation of experimental data
it is important to have a clear and correct
theoretical formulation
of the neutrino oscillation mechanism\footnote{
See Ref.~\cite{Beuthe:2001rc} for a review
of the theory of neutrino oscillations
and Ref.~\cite{Neutrino_Unbound}
for an exhaustive list of references.
}.
In particular,
the phase $\Phi_{kj}$ of neutrino oscillations in vacuum
due to the interference of the contributions of two massive neutrinos
$\nu_k$ and $\nu_j$
with masses
$m_k$ and $m_j$,
respectively,
depends on the
mass-squared differences
$\Delta{m}^2_{kj} \equiv m_k^2 - m_j^2$.
A correct theoretical expression of the
oscillation phase $\Phi_{kj}$
is important in order to extract from the experimental data
correct information on the value of
$\Delta{m}^2_{kj}$.

The standard expression for the oscillation phase
$\Phi_{kj}$
in the relativistic approximation is
\begin{equation}
\Phi_{kj}
=
- \frac{\Delta{m}^2_{kj} \, L}{2 \, E}
\,,
\label{011}
\end{equation}
where $E$ is the neutrino energy
and $L$ is the distance between the neutrino source and detector.

It has been argued by some authors
\cite{Rotelli-99,Field:2001xf}
that the expression (\ref{011}) is wrong by a factor of two
on the basis of the following reasoning.
Neutrino experiments measure neutrino oscillations as a function of
the source-detector distance $L$.
Different massive neutrinos
propagate with different velocities
\begin{equation}
v_k = \frac{p_k}{E_k}
\,,
\label{012}
\end{equation}
where $E_k$ and $p_k$
are, respectively, the energy and momentum of the neutrino with mass $m_k$,
related by the relativistic dispersion relation
\begin{equation}
E_k^2 = p_k^2 + m_k^2
\,.
\label{0121}
\end{equation}
Hence,
the phases of the different massive neutrinos
wave functions after a propagation distance $L$
should take into account the different times of propagation of different massive neutrinos:
\begin{equation}
\widetilde\Phi_k
=
p_k \, L
- E_k \, t_k
\,.
\label{013}
\end{equation}
The different propagation times are given by
\begin{equation}
t_k
=
\frac{L}{v_k}
=
\frac{E_k}{p_k} \, L
\,,
\label{014}
\end{equation}
which lead, in the relativistic approximation, to the phase difference
\begin{equation}
\Delta\widetilde\Phi_{kj}
\equiv
\widetilde\Phi_k - \widetilde\Phi_j
=
- \frac{\Delta{m}^2_{kj} \, L }{ E }
\,.
\label{015}
\end{equation}
This phase difference is twice of the standard one in Eq.~(\ref{011}).
A similar disagreement by a factor of two
has been claimed to exist in the case of kaon oscillations
in Refs.~\cite{Srivastava-Widom-Sassaroli-Lambda-95,%
Srivastava-Widom-Sassaroli-correlations-95,%
Srivastava-Widom-Lambda-96}.

Let us notice that in Eq.~(\ref{013})
we have considered the possibility of different energies and momenta
for different massive neutrino wave functions.
Indeed,
it has been shown in Ref.~\cite{Giunti:2001kj}
that Lorentz invariance implies that
in general in oscillation experiments
different massive neutrinos have different energies
and different momenta.
The energy $E$ in Eqs.~(\ref{011}) and (\ref{015})
is the energy of the massive neutrinos in the massless limit,
\textit{i.e.}
neglecting the differences of energy due to the masses.

The authors of Refs.~\cite{Lipkin:1995cb,%
Grossman:1997eh,%
Lipkin:1999nb}
claimed that a correct way to obtain the standard oscillation phase
is to assume the same energy for the different massive neutrino wave functions.
This is an unphysical assumption,
as discussed in Ref.~\cite{Giunti:2001kj}.
Moreover,
as already noticed in Ref.~\cite{Rotelli-99},
it is not true that the disagreement of a factor of two disappears
assuming the same energy for the different massive neutrino wave functions,
as clearly shown by the above derivation of Eq.~(\ref{015}),
in which the energies of the different massive neutrino wave functions
could have been taken to be equal.
Indeed,
even if the different massive neutrino wave functions
have the same energy,
the time contribution
$-E t_k+E t_j$
to the phase difference $\Delta\widetilde\Phi_{kj}$
does not disappear,
because $t_k \neq t_j$.
This contribution has been missed in Refs.~\cite{Lipkin:1995cb,%
Grossman:1997eh,%
Lipkin:1999nb}.

\begin{figure}
\begin{center}
\includegraphics[bb=90 468 526 772, width=0.8\textwidth]{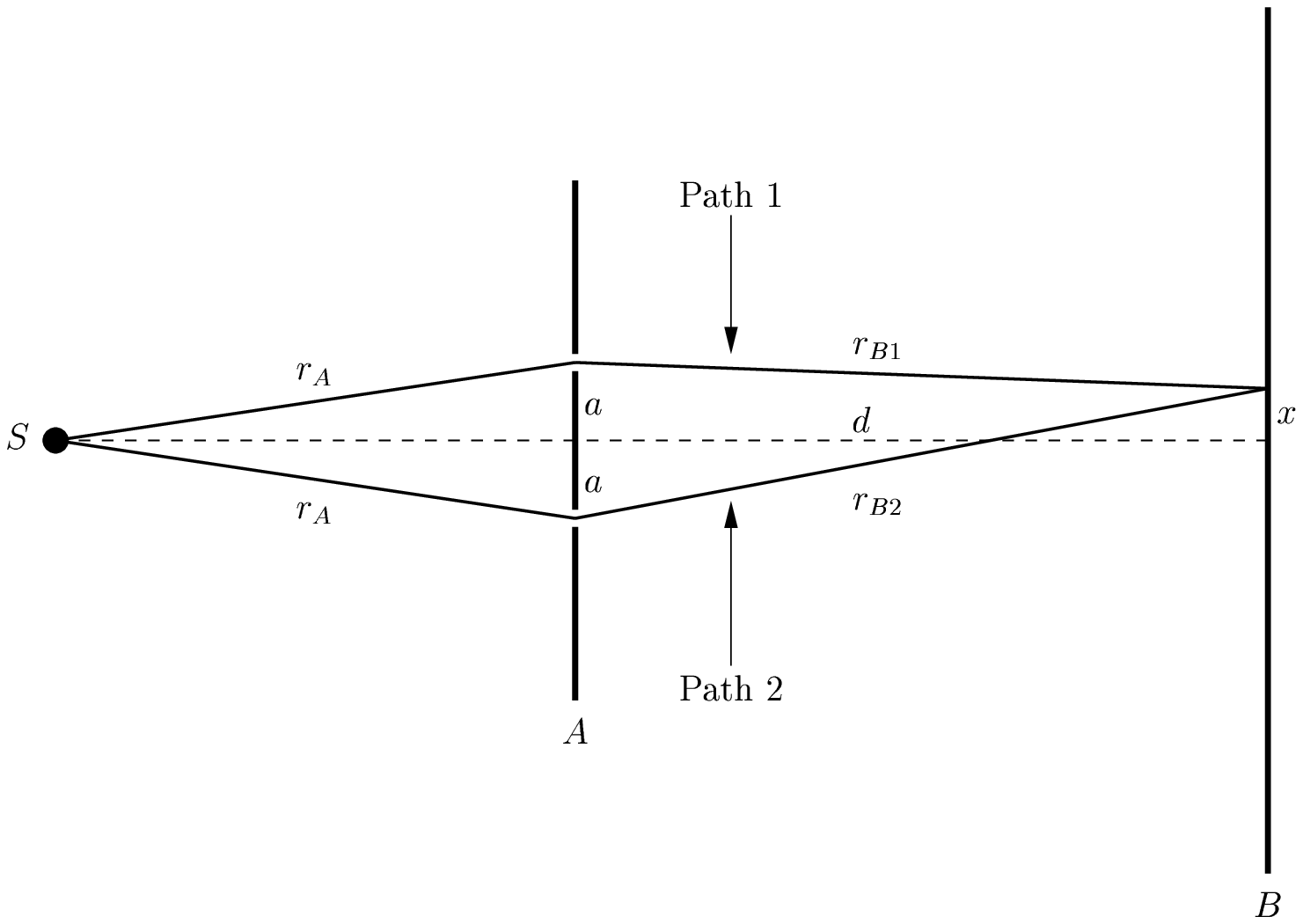}
\end{center}
\caption{ \label{young}
Schematic view of Young double-slit interference experiment.
}
\end{figure}

In order to test the validity of the
reasoning of the authors of Refs.~\cite{Rotelli-99,Field:2001xf},
let us apply their method to the well-known
double-slit interference experiment
depicted in Fig.~\ref{young}.
The particles emitted by the source $S$
and detected  could be
photons or electrons, or others.
The screen $A$ has two holes through which the particles can reach the
screen $B$ on which a detector registers the arrival
of the particles at a distance $x$ from the center of the screen.

In the standard approach,
the phases of the two waves at $x$ at the time $t$ are given by
\begin{equation}
\Phi_k
=
p \, r_k - E \, t
=
p \left( r_A + r_{Bk }\right) - E \, t
\,,
\label{001}
\end{equation}
for $k=1,2$.
Hence, the phase difference is
\begin{equation}
\Delta{\Phi} \equiv \Phi_2 - \Phi_1
=
p \, \Delta{r}
\,,
\label{002}
\end{equation}
with
\begin{equation}
\Delta{r} \equiv r_2 - r_1 = r_{B2} - r_{B1} \simeq
\frac{2 \, a \, x}{d}
\,,
\label{026}
\end{equation}
where the approximation holds for
$x,a \ll d$.
Since $p = 2\pi / \lambda$,
where $\lambda$ is the wavelength,
the maxima of interference are given by the usual well known formula
$x = n \lambda d / 2 a$,
with $n=0,1,\ldots$,
which has been confirmed by many experiments without any doubt.

Let us apply now the reasoning of the authors of Refs.~\cite{Rotelli-99,Field:2001xf}.
If one takes into account the velocity of the particles,
the two paths have different propagation times
\begin{equation}
t_k = \frac{r_k}{v}
\,,
\label{004}
\end{equation}
where $v$ is the velocity,
\begin{equation}
v = \frac{p}{E}
\,.
\label{005}
\end{equation}
In this approach,
the phases of the two waves at $x$ are given by
\begin{equation}
\widetilde\Phi_k = p \, r_k - E \, t_k
\,.
\label{003}
\end{equation}
The phase difference at $x$
turns out to be
\begin{equation}
\Delta\widetilde\Phi
\equiv
\widetilde\Phi_2 - \widetilde\Phi_1
=
\left( p - \frac{E}{v} \right) \Delta{r}
\simeq
- 2 \, \frac{a}{d} \, x \, \frac{m^2}{p}
\,,
\label{006}
\end{equation}
where the approximation holds for
$x,a \ll d$.
This phase difference is very different from the correct one
in Eq.~(\ref{002}),
and it even vanishes for massless particles
(as photons).

Therefore,
the reasoning presented in Refs.~\cite{Rotelli-99,Field:2001xf}
is wrong.

The mistake in the claim formulated in Refs.~\cite{Rotelli-99,Field:2001xf}
is due to a wrong use of the group velocity
in the phase,
which depends on the phase velocity.
The group velocity
(given in Eq.~(\ref{012}) for neutrinos and in Eq.~(\ref{005})
for the double-slit experiment)
is the velocity of the factor that modulates the amplitude
of a wave packet describing a localized particle.
In the double-slit experiment there are two wave packets
which propagate along the two paths in Fig.~\ref{young}.
The envelopes of these wave packets take different times
to cover the two different distances from
the source $S$ to the point $x$ on the screen.
But this has no effect on the phases.
Only the amplitude of the final wave function is
determined by the modulating factors of the amplitudes of the
two wave packets.
The different arrival times of the envelopes of the wave packets
reduces the overlap of the two wave packets,
leading to a decoherence effect.

\begin{figure}
\begin{center}
\includegraphics[bb=100 437 418 755, width=0.8\textwidth]{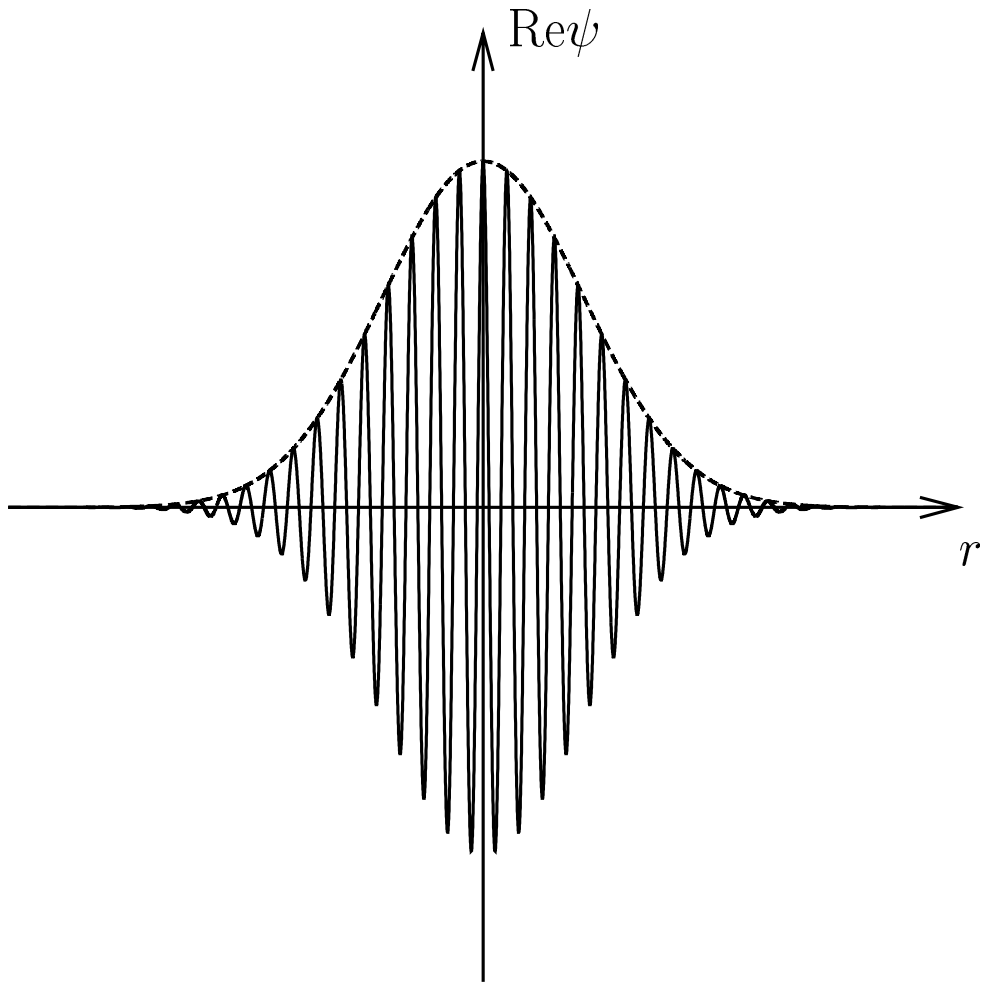}
\end{center}
\caption{ \label{wp}
Real part of the wave packet (\ref{021}) at $t=0$.
The dashed line represents the envelope (\ref{022}).
}
\end{figure}

Let us illustrate these concepts in a one-dimensional formalism
using a Gaussian momentum distribution
with width $\sigma_p$,
centered at the momentum $p$:
\begin{equation}
\psi(p')
=
\left( \sqrt{2\pi} \, \sigma_p \right)^{-1/2}
\exp\left[
-
\frac{ \left( p' - p \right)^2 }{ 4 \sigma_p^2 }
\right]
\,.
\label{020}
\end{equation}
For
$\sigma_p << p$,
the corresponding wave packet in space-time is
\begin{equation}
\psi(r,t)
=
\left( \sqrt{2\pi} \, \sigma_r \right)^{-1/2}
\exp\left[
i \, p \, r
-
i \, E \, t
-
\frac{ \left( r - v t \right)^2 }{ 4 \sigma_r^2 }
\right]
\,,
\label{021}
\end{equation}
where $\sigma_r=1/2\sigma_p$, $E=\sqrt{p^2+m^2}$, and $v=p/E$
is the velocity of the envelope
\begin{equation}
\left( \sqrt{2\pi} \, \sigma_r \right)^{-1/2}
\exp\left[
-
\frac{ \left( r - v t \right)^2 }{ 4 \sigma_r^2 }
\right]
\label{022}
\end{equation}
of the wave packet.
The real part of the wave packet (\ref{021}) at $t=0$
is depicted by the solid line in Fig.~\ref{wp},
with the envelope (\ref{022}) represented by the dashed line.
It is important to understand that the
wave packet in the approximation (\ref{021}) is a monochromatic wave
whose amplitude is modulated by the envelope factor (\ref{022}),
that has no effect on the phase factor
\begin{equation}
\exp\left[
i \, p \, r
-
i \, E \, t
\right]
\label{023}
\end{equation}
of the wave packet.

The amplitudes of each of
the two wave packets
traveling the two paths in the double-slit experiment in Fig.~\ref{young}
at the distance $x$ from the center of screen $B$ at the time $t$ is given by
\begin{equation}
\psi(r_k,t)
=
\frac{1}{\sqrt{2}}
\left( \sqrt{2\pi} \, \sigma_r \right)^{-1/2}
\exp\left[
i \, p \, r_k
-
i \, E \, t
-
\frac{ \left( r_k - v t \right)^2 }{ 4 \sigma_r^2 }
\right]
\,,
\label{024}
\end{equation}
for
$k=1,2$,
with
$r_k = r_A + r_{Bk}$.
The probability to detect the particle at the point $x$ at the time $t$ is
\begin{eqnarray}
P(x,t)
&\propto&
\left|
\psi(r_1,t)
+
\psi(r_2,t)
\right|^2
\nonumber
\\
&\propto&
\frac{1}{2 \, \sqrt{2\pi} \, \sigma_r}
\left\{
\exp\left[
-
\frac{ \left( r_1 - v t \right)^2 }{ 2 \sigma_r^2 }
\right]
+
\exp\left[
-
\frac{ \left( r_2 - v t \right)^2 }{ 2 \sigma_r^2 }
\right]
\right.
\nonumber
\\
&&
\hspace{2cm}
\left.
+ 2 \, \cos\!\left(p\Delta{r}\right)
\exp\left[
-
\frac{ \left( r_1 - v t \right)^2 + \left( r_2 - v t \right)^2 }{ 4 \sigma_r^2 }
\right]
\right\}
\,,
\label{025}
\end{eqnarray}
where we have omitted an appropriate normalization factor,
that must be inserted in order to normalize the probability
over the screen.

From Eq.~(\ref{025})
it is clear that the probability to detect the particle at $x$
is appreciable only when one of the two wave packets
cover the detector.
Interference of the two wave functions
is possible if both wave packets cover the detector
simultaneously.
However,
usually the detector operates for a long time
without recording the precise instant of detection.
In this case,
the probability of detection at $x$ is given by the time average
of $P(x,t)$ in Eq.~(\ref{025}):
\begin{equation}
P(x)
\propto
\frac{1}{2}
\left\{
1
+
\cos\!\left(p\Delta{r}\right)
\exp\left[
-
\frac{ \Delta{r}^2 }{ 8 \sigma_r^2 }
\right]
\right\}
\,.
\label{027}
\end{equation}
The interference term
gives an appreciable contribution only if
the separation $\Delta{r}$ of the two wave packets is smaller than their width,
\textit{i.e.} if
there is an overlap of the two wave packets at the detector.
It is clear that the different propagation times of the envelopes
of the two wave packets
has no effect on the phase of the interference term,
that is equal to the phase in Eq.~(\ref{002}),
obtained in the plane wave approximation.
This is what we wanted to show with this simple example.

An interesting complication to the problem
stems from the fact,
noted in Ref.~\cite{Kiers-Nussinov-Weiss-PRD53-96},
that different wave packets
can interfere even if they do not overlap at the detector
if the coherence time of the detection process
is larger than the time separation between the arrival of the wave packets.
In order to take into account this effect
in the double-slit experiment,
we have to introduce an appropriate wave function for the detection process.
Assuming that the detection process has a coherence time $\sigma_{t\mathrm{D}}$
and the detection occurs at the time $t_{\mathrm{D}}$,
we describe the detection process with the Gaussian wave packet
\begin{equation}
\psi_{\mathrm{D}}(t,t_{\mathrm{D}})
=
\left( \sqrt{2\pi} \, \sigma_{t\mathrm{D}} \right)^{-1/2}
\exp\left[
- i \, E \left( t - t_{\mathrm{D}} \right)
-
\frac{ \left( t - t_{\mathrm{D}} \right)^2 }{ 4 \sigma_{t\mathrm{D}}^2 }
\right]
\,.
\label{031}
\end{equation}
The frequency of the detection process
is the same as that of the absorbed wave function,
because the incoming wave function
excites the corresponding degree of freedom
of the detection process.
The amplitude of detection of each of the two incoming wave packets
is given by the overlap of the incoming wave packet
(\ref{024})
and the detection process wave packet
(\ref{031}):
\begin{equation}
A(r_k,t_{\mathrm{D}})
\propto
\int \mathrm{d}t
\,
\psi_{\mathrm{D}}^{*}(t,t_{\mathrm{D}})
\,
\psi(r_k,t)
\,.
\label{032}
\end{equation}
Performing the integration over $t$,
we obtain
\begin{equation}
A(r_k,t_{\mathrm{D}})
\propto
\frac{1}{\sqrt{2}}
\left( \sqrt{2\pi} \, \sigma \right)^{-1/2}
\,
\exp\left[
i \, p \, r_k
-
i \, E \, t_{\mathrm{D}}
-
\frac{ \left( r_k - v t_{\mathrm{D}} \right)^2 }{ 4 \sigma^2 }
\right]
\,,
\label{033}
\end{equation}
with
\begin{equation}
\sigma^2 \equiv \sigma_r^2 + v^2 \, \sigma_{t\mathrm{D}}^2
\,.
\label{034}
\end{equation}
One can see that the time allowed for a coherent absorption of
the incoming wave packet is increased by the coherence time of
the detection process,
that dominates if $v \, \sigma_{t\mathrm{D}} \gg \sigma_r$.

The probability to detect the particle at $x$ at the time $t_{\mathrm{D}}$
is now given by
\begin{eqnarray}
P(x,t_{\mathrm{D}})
&\propto&
\left|
A(r_1,t_{\mathrm{D}})
+
A(r_2,t_{\mathrm{D}})
\right|^2
\nonumber
\\
&\propto&
\frac{1}{2 \, \sqrt{2\pi} \, \sigma}
\left\{
\exp\left[
-
\frac{ \left( r_1 - v t_{\mathrm{D}} \right)^2 }{ 2 \sigma^2 }
\right]
+
\exp\left[
-
\frac{ \left( r_2 - v t_{\mathrm{D}} \right)^2 }{ 2 \sigma^2 }
\right]
\right.
\nonumber
\\
&&
\hspace{2cm}
\left.
+ 2 \, \cos\!\left(p\Delta{r}\right)
\exp\left[
-
\frac{ \left( r_1 - v t_{\mathrm{D}} \right)^2 + \left( r_2 - v t_{\mathrm{D}} \right)^2 }{ 4 \sigma^2 }
\right]
\right\}
\,.
\label{035}
\end{eqnarray}
This equation has the same structure as Eq.~(\ref{025}),
with the width $\sigma_r$ of the wave packets
replaced by $\sigma$.
Obviously,
the average over the unmeasured detection time $t_{\mathrm{D}}$
leads to an expression for the probability to detect the particle at $x$
given by Eq.~(\ref{027}) with $\sigma_r$ replaced by $\sigma$,
which takes into account also the coherence time of
the detection process.

Therefore, we have seen that the coherence time of
the detection process,
which allows a coherent absorption of wave packets
arriving at the detector at different times,
do not have any effect on the phase of the interference
between different wave functions,
refuting the arguments presented in Refs.~\cite{Rotelli-99,Field:2001xf}.
The resulting prescription
in calculations performed in the plane wave approximation
(\textit{i.e} neglecting the wave packet character of wave functions
that describe localized particles)
is to calculate the interference of different wave functions
at the same time and the same space point
\cite{Lowe-Lambda-96,Kayser-QM-97}.

The physical explanation of the independence of the phase of the interference
between different wave functions
from their arrival time at the detector
has been presented in Ref.~\cite{Giunti:2000kw}.
It consists in taking into account that there is interference only
if the different wave functions are detected coherently.
Coherence means that
there is a precise phase difference of the detection process
between the arrival times of the different wave functions
that
must be taken into account.
When the second wave packet arrives, the detection process
is already excited with a frequency due to its interaction with the first
wave packet and its phase is determined by the time difference
between the arrivals of the two wave packets.
It is easy to see
\cite{Giunti:2000kw}
that the phase of the detection process cancels exactly the
phase difference of the two wave functions due to the different detecting times,
leading to the practical prescription
to calculate the interference of different wave functions
at the same time and the same space point.

Let us now apply the wave packet formalism to neutrino oscillations,
following the lines presented in Refs.~\cite{Giunti:1991ca,Giunti:1998wq}.
The Gaussian wave packets that describe massive neutrinos in one spatial dimension $x$
along the source-detector direction
are
\begin{equation}
\psi_k(x,t)
=
\left( \sqrt{2\pi} \, \sigma_{x\mathrm{P}} \right)^{-1/2}
\exp\left[
i \, p_k \, x
-
i \, E_k \, t
-
\frac{ \left( x - v_k t \right)^2 }{ 4 \sigma_{x\mathrm{P}}^2 }
\right]
\,,
\label{041}
\end{equation}
where $\sigma_{x\mathrm{P}}$ is the width of the wave packets,
that is determined by the production process,
$E_k$ and $p_k$
are, respectively, the energy and momentum of the neutrino with mass $m_k$,
related by the relativistic dispersion relation (\ref{0121}),
and
$v_k$ is the group velocity given in Eq.~(\ref{012}).

Taking into account the possibility that the detection process
has a finite coherence time interval $\sigma_{t\mathrm{D}}$
and a finite spatial coherence width $\sigma_{x\mathrm{D}}$,
we describe the detection process of the massive neutrino $\nu_k$
with the Gaussian wave packet
\begin{equation}
\psi_{\mathrm{D}k}(x,t,L,T)
\propto
\exp\left[
i \, p_k \left( x - L \right)
- i \, E_k \left( t - T \right)
-
\frac{ \left( x - L \right)^2 }{ 4 \sigma_{x\mathrm{D}}^2 }
-
\frac{ \left( t - T \right)^2 }{ 4 \sigma_{t\mathrm{D}}^2 }
\right]
\,,
\label{042}
\end{equation}
where $L$ is the source-detector distance and $T$ is the time elapsed
between neutrino production and detection.
Notice that
with respect to the wave packet (\ref{031}),
in Eq.~(\ref{042})
we have considered also a finite spatial coherence width
of the detection process that was neglected
in the discussion of the double-slit experiment.
On the other hand, with respect to the wave packet treatment
of neutrino oscillations presented in Ref.~\cite{Giunti:1998wq}
we have added the explicit terms
that take into account the
finite coherence time interval $\sigma_{t\mathrm{D}}$
of the detection process.

The frequency and wave number of the detection process
are the same as those of the detected massive neutrino wave function,
which excites the corresponding degree of freedom
of the detection process.
In the framework of quantum field theory
the detection process is described in terms of the leptonic weak charged current
\begin{equation}
j^\mu(x)
=
\sum_{\alpha=e,\mu,\tau}
\sum_k
\overline{\alpha}(x)
\,
\gamma^\mu
\left( 1 - \gamma^5 \right)
\,
U_{\alpha k}
\,
\nu_k(x)
\,,
\label{0421}
\end{equation}
where $U$ is the lepton mixing matrix
(see \cite{Bilenky-Pontecorvo-PR-78,%
Bilenky-Petcov-RMP-87,%
CWKim-book-93,%
BGG-review-98}).
Only the field $\nu_k(x)$
is excited by the arrival of the wave function of the
corresponding massive neutrino $\nu_k$,
and the frequency and wave number of the field excitations
are the same as those of the incoming wave function.

The amplitude of detection of each massive neutrino wave packet
is given by the overlap of the incoming neutrino wave packet
with the detection process wave packet:
\begin{equation}
A_k(L,T)
\propto
\int \mathrm{d}x \, \mathrm{d}t \,
\psi_{\mathrm{D}k}^{*}(x,t,L,T)
\,
\psi_k(x,t)
\,.
\label{043}
\end{equation}
Performing the integration over $x$ and $t$ we obtain
\begin{equation}
A_k(L,T)
\propto
\exp\left[
i \, p_k \, L
-
i \, E_k \, T
-
\frac{ \left( L - v_k T \right)^2 }{ 4 \sigma_{xk}^2 }
\right]
\,,
\label{044}
\end{equation}
with
\begin{equation}
\sigma_{xk}^2
\equiv
\sigma_{x\mathrm{P}}^2 + \sigma_{x\mathrm{D}}^2 + v_k^2 \, \sigma_{t\mathrm{D}}^2
\,.
\label{045}
\end{equation}
This result is simple and important.
It shows clearly that the coherent absorption of each massive neutrino wave function
occurs in a space-time interval of width $\sigma_{xk}$
around the coordinates $L,T$ of the detection process.
The size of this space-time interval
is determined by the width of the neutrino wave packet
and the coherence spatial and temporal widths
of the detection process.
The largest width dominates.
The contribution of the temporal width of the detection process
is weighted by the velocity of the incoming neutrino wave packet for obvious reasons.
Since this velocity depends on the mass of the neutrinos,
the width $\sigma_{xk}$
depends on the index $k$ labeling massive neutrinos.
However,
for relativistic neutrinos the contribution of neutrino mass to
$\sigma_{xk}$
is negligible
and we can safely approximate
\begin{equation}
\sigma_{xk}^2
\simeq
\sigma_{x\mathrm{P}}^2 + \sigma_{x\mathrm{D}}^2 + \sigma_{t\mathrm{D}}^2
\equiv
\sigma_{x}^2
\,.
\label{046}
\end{equation}
Notice that, on the other hand,
the contribution of neutrino mass to the other terms in Eq.~(\ref{044})
cannot be neglected because it is amplified by
the macroscopic quantities $L$ and $T$.
In the following we use the relativistic approximation,
which implies that
\begin{equation}
E_k
\simeq
E
+
\xi
\,
\frac{ m_{k}^2 }{ 2 E }
\,,
\qquad
p_k
\simeq
E
-
\left( 1 - \xi \right)
\frac{ m_{k}^2 }{ 2 E }
\,,
\qquad
v_k
\simeq
1 - \frac{ m_{k}^2 }{ 2 E^2 }
\,,
\label{0461}
\end{equation}
where $E$ is the neutrino energy in the limit of zero mass.
The first order correction to the momentum and energy
due to neutrino mass $m_k$ is proportional to $m_{k}^2$
because of the relativistic dispersion relation
(\ref{0121}),
and must be divided by the energy $E$
for dimensional reasons.
The coefficient $\xi$
depends on the production process,
but we will see that it has no effect on the phase of neutrino oscillations,
as already shown in Refs.~\cite{Giunti:1991ca,Giunti:1998wq,Giunti:2000kw}.

The probability of flavor transitions is given by
(see \cite{Bilenky-Pontecorvo-PR-78,%
Bilenky-Petcov-RMP-87,%
CWKim-book-93,%
BGG-review-98})
\begin{equation}
P_{\nu_\alpha\to\nu_\beta}(L,T)
\propto
\left|
\sum_k
U_{{\alpha}k}^{*} \, A_k(L,T) \, U_{{\beta}k}
\right|^2
\,,
\label{047}
\end{equation}
where $U$ is the lepton mixing matrix.
Performing the average of $P_{\nu_\alpha\to\nu_\beta}(L,T)$ over the
unmeasured neutrino
propagation time $T$,
we finally obtain
\begin{equation}
P_{\nu_\alpha\to\nu_\beta}(L)
=
\sum_{k}
|U_{{\alpha}k}|^2 \, |U_{{\beta}k}|^2
+
2 \, \mathrm{Re}
\sum_{k>j}
U_{{\alpha}k}^{*}
\,
U_{{\beta}k}
\,
U_{{\alpha}j}
\,
U_{{\beta}j}^{*}
\,
F_{kj}
\,
\exp\left[
- i \, \frac{ \Delta{m}^2_{kj} \, L }{ 2 E }
- \left( \frac{L}{L_{kj}^{\mathrm{coh}}} \right)^2
\right]
\,,
\label{048}
\end{equation}
where
\begin{equation}
L_{kj}^{\mathrm{coh}}
=
\frac{4 \, \sqrt{2} \, E^2}{|\Delta{m}^2_{kj}|} \, \sigma_x
\label{049}
\end{equation}
are the coherence lengths and
\begin{equation}
F_{kj}
=
\exp\left[
- 2 \pi^2 \, \xi^2 \,
\left( \frac{\sigma_x}{L_{kj}^{\mathrm{osc}}} \right)^2
\right]
\,,
\label{050}
\end{equation}
with the oscillation lengths
\begin{equation}
L_{kj}^{\mathrm{osc}}
=
\frac{4 \pi \, E}{|\Delta{m}^2_{kj}|}
\,.
\label{051}
\end{equation}
The coherence length
$L_{kj}^{\mathrm{coh}}$
is the distance beyond which the contributions of the
massive neutrinos $\nu_k$ and $\nu_j$
do not interfere any more
because the separation between their wave packets
is larger than
the size of the wave packets
and the spatial and temporal coherence widths of
the detection process.
The coefficient $F_{kj}$
suppresses the interference of $\nu_k$ and $\nu_j$ if
$\sigma_x \gtrsim L_{kj}^{\mathrm{osc}}$,
\textit{i.e.}
if the production or the detection process
is not localized in a space-time region much smaller than the
oscillation length.
This obvious constraint for the observation of neutrino oscillations
was discussed for the first time in Ref.~\cite{Kayser-oscillations-81}
and is satisfied in all experiments.
Therefore,
in practice one can safely approximate $F_{kj} \simeq 1$.

From Eq.~(\ref{048})
one can see that a proper wave packet treatment of neutrino oscillations
that takes into account the different propagation times
of massive neutrino wave packets
leads to the standard oscillation phase (\ref{011}),
refuting the disagreement of a factor of two claimed in
Refs.\cite{Rotelli-99,Field:2001xf}.
Neutrino oscillations are due to the different phase velocities
of different massive neutrinos,
which produces interference.
This interference is obviously the same
whether it is calculated in the plane wave approach or
in the wave packet treatment.
The different conclusion reached in Refs.\cite{Rotelli-99,Field:2001xf}
is due to a wrong use of the group velocity
in the phase.

In conclusion,
using an analogy with the well-known double-slit experiment
we have shown that the standard phase (\ref{011}) of neutrino oscillations
is correct,
refuting the claim of a factor of two correction presented in
Refs.\cite{Rotelli-99,Field:2001xf}.
We have also improved the wave packet treatment of neutrino oscillations
presented in Refs.~\cite{Giunti:1991ca,Giunti:1998wq},
taking into account explicitly
the finite coherence time of the detection process.

\clearpage

\end{document}